\documentclass[]{aa}

\usepackage{amsmath}
\usepackage{amssymb}
\usepackage{graphicx}
\usepackage{txfonts}
\usepackage{natbib}
\bibpunct{(}{)}{;}{a}{}{,} 

\newcommand\cm{\,\rm cm}
\newcommand\cmt{\,\rm cm^{-3}}
\newcommand\s{\,\rm s}

\newcommand\erg{\,\rm erg}
\newcommand\K{\,\rm K}

\newcommand\Myr{\,\rm Myr}

\newcommand\pc{\,\rm\,pc}

\newcommand\lF{\lambda_F}

\usepackage{ulem}

\begin{document}

  \title{A {F}ield-length based refinement criterion for
         adaptive mesh simulations of the interstellar medium}
  \titlerunning{A {F}ield-length criterion for {AMR} simulations}
  \author{Oliver Gressel\inst{1,2}}
  \offprints{O. Gressel\\\email{o.gressel@qmul.ac.uk}}
  \institute{Astrophysikalisches Institut Potsdam, 
             An der Sternwarte 16, D-14482 Potsdam, Germany \and
             Astronomy Unit, School of Mathematical Sciences,
             Queen Mary, University of London, Mile End Road, London E1~4NS, UK}
  \date{}

  \abstract
    {Adequate modelling of the multiphase interstellar medium requires
     optically thin radiative cooling, comprising an inherent thermal
     instability. The size of the occurring condensation and evaporation
     interfaces is determined by the so-called Field-length, which gives the
     dimension at which the instability is significantly damped by thermal
     conduction.}
    {Our aim is to study the relevance of conduction scale effects in
     the numerical modelling of a bistable medium and check the
     applicability of conventional and alternative adaptive mesh
     techniques.}
    {The low physical value of the thermal conduction within the ISM defines a
     multiscale problem, hence promoting the use of adaptive meshes. We here
     introduce a new refinement strategy that applies the Field condition by
     Koyama \& Inutsuka as a refinement criterion. The described method is
     very similar to the Jeans criterion for gravitational instability by
     Truelove and efficiently allows to trace the unstable gas situated at the
     thermal interfaces.}
    {We present test computations that demonstrate the greater accuracy of the
     newly proposed refinement criterion in comparison to refinement based on
     the local density gradient. Apart from its usefulness as a refinement
     trigger, we do not find evidence in favour of the Field criterion as a
     prerequisite for numerical stability.}
    {}
   \maketitle


\section{Introduction} \label{sec:intro}

The structure of the turbulent interstellar medium (ISM) of the Milky
Way and other disk galaxies is determined by various complex physical
processes. In particular, the formation of the neutral atomic phase of
the ISM is believed to be regulated by condensation under the action
of thermally bistable radiative cooling. The related thermal
instability (TI) has been extensively studied as a modifying agent for
various driving sources of turbulence in the ISM. Numerical studies
include the magneto-rotational instability (MRI)
\citep{2005ApJ...629..849P,2007ApJ...663..183P}, explosions of SNe
\citep{1999ApJ...514L..99K,2005A&A...436..585D,%
2006ApJ...653.1266J,2008AN....329..619G,2008A&A...486L..35G}, and
converging flows \citep{2005A&A...433....1A,2006ApJ...643..245V,%
2008ApJ...683..786H}. Adaptive mesh MHD simulations of the last
scenario have been performed by \citet{2008A&A...486L..43H}, who
applied a simple density threshold as a refinement criterion.

The development of TI alone has been explored under various conditions
by \citep{2002ApJ...577..768S}. Recently, there has been a discussion
whether TI in conjunction with thermal conduction can act as an
independent source of turbulence
\citep{2007ApJ...654..945B,2006astro.ph..5528K}.

In dynamic simulations of the ISM, the radiative cooling is usually
treated in the limit of an optically thin plasma, i.e., radiative
transport effects are neglected. The cooling rate is then prescribed
by $\Lambda(T)$ in units of $\erg\,\cm^3s^{-1}g^{-2}$. Here we adopt
the cooling function from \cite{2002ApJ...577..768S} that is a
piecewise power-law fit to results by \cite{1995ApJ...443..152W}.

In this paper, we want to focus on an issue in the numerical treatment
of thermally bistable gas that has firstly been brought up by
\cite{2004ApJ...602L..25K}: The comprehensive theoretical analysis by
\cite{1965ApJ...142..531F} reveals that, in the case of vanishing
thermal conduction, TI has a finite growth rate in the limit of high
wave numbers. This implies that, in the discrete representation of the
fluid, numerical fluctuations at grid scale will be prone to
artificial amplification through the physical instability unless one
does take suitable precautions. The most natural way to guarantee a
physically meaningful, converged solution is to introduce a physically
motivated dissipation length that then needs to be resolved on the
numerical grid. For the case when thermal conduction cannot be
neglected, \cite{1965ApJ...142..531F} derives a stability criterion
for the condensation mode which has the general form
$\,d(\rho\Lambda)/dT - \rho/T\,d(\rho\Lambda)/d\rho < -k^2\,\kappa\,$
with $\kappa$ the coefficient of thermal conduction in units of
$\erg\,\cm^{-1}\K^{-1}\s^{-1}$. In the case of a power-law cooling
function $\Lambda(T)\sim T^{\beta}$ this reduces to
$\,(\beta\!-\!1)\,\rho^2\Lambda/T < -k^2\,\kappa\,$. The wavelength
$\lF$ where the criterion is marginally fulfilled marks the scale
where thermal conduction efficiently damps out fluctuations generated
by the TI. \cite{2004ApJ...602L..25K} have shown that
\begin{equation}
  \lF=\frac{2\pi}{\sqrt{1-\beta}}\,\sqrt{\frac{\kappa T}{\rho^2\Lambda}}
  \label{eq:lF}
\end{equation}
has to be resolved with at least three grid cells to attain a
converged numerical solution. It occurs naturally to use this
criterion as a condition for mesh refinement.


\section{Numerical Methods} \label{sec:methods}

For our simulations, we make use of the newly developed version 3 of
the NIRVANA code which is a general purpose MHD fluid tool designed to
solve multiscale, self-gravitating magnetohydrodynamics problems. The
NIRVANA code comprises (i) a fully conservative, divergence-free
Godunov-type central scheme, \citet{2004JCoPh.196..393Z}, (ii) block
structured adaptive mesh refinement (AMR), and (iii) a multigrid-like
adaptive mesh Poisson solver with elliptic matching,
\citet{2005CoPhC.170..153Z}.

The central aim of adaptive mesh simulations is to resolve regions of
interest at enhanced accuracy. In AMR codes like Enzo
\citep{2004astro.ph..3044O}, FLASH \citep{2000ApJS..131..273F},
NIRVANA, or RAMSES \citep{2002A&A...385..337T}, this can be done by a
variety of different criteria: baryon or dark matter overdensity
(cosmological structure formation), local Jeans-length (protostellar
core collapse), or entropy gradients (accretion shocks). Recently,
\citet{2008MNRAS.388.1079I} have introduced a vorticity-based
criterion to properly resolve a turbulent wake. Alternatively,
independent of the underlying physics, the amount of discrepancy
between the solutions on different AMR levels can be used as an
indication for refinement \citep{1989JCoPh..82...64B}.

\subsection{Classical mesh refinement} \label{sec:classical}

As a standard criterion to trace developing structures, one typically
monitors local gradients $\delta u$ in the fluid variables $u$. To be
able to follow moving structures (and to anticipate emerging
structures), it is desirable to furthermore consider the temporal
change of these gradients. Due to the additional overhead in storing
quantities between the timesteps, this is usually discarded. Instead,
one can make use of the fact that strong gradients are bordered by
sections of enhanced curvature -- reflected in the second derivatives
$\delta^2u$. Accordingly, NIRVANA's usual mesh refinement is
determined by combining the criterion of normalised gradients
$\|\delta u\|_2:|u|$ with one of the ratio of second to first
derivatives $\|\delta^2u\|_2:\|\delta u\|_2$ of the conserved
variables $u$, i.e.,
\begin{equation}
\left[\,\alpha\,\frac{\|\delta u\|_2}{|u|}
     +(1\!-\!\alpha)\,\frac{\|\delta^2u\|_2}
     {\|\delta u\|_2+\epsilon\,|u|}\,\right]
     \ \left(\frac{\Delta x^{(l)}}{\Delta x^{(0)}}\right)^{\xi}
     \begin{cases} > \epsilon_u       & \exists u \\
                   < 0.7 \epsilon_u & \forall u \end{cases}
     \label{eq:classical}
\end{equation}
where $\|\delta u\|_2$ and $\|\delta^2 u\|_2$ are the first and second
numerical derivatives in the Euclidian norm, and $\alpha \in [0,1]$
controls the mixture of the two indicators. A new block is created if
the criterion exceeds a critical value $\epsilon_u$ in a two-cell wide
buffer zone around the block to be inserted. Blocks are removed if the
criterion falls below a factor $0.7$ times the critical value. By
tuning $\alpha$, the user can specify a bias towards the first or
second type of criterion; a value of $\alpha=0$ corresponds to the
standard estimator used in the FLASH code \citep{2000ApJS..131..273F}.

For our runs not based on the Field condition, we adopt refinement for
the gas density $\rho$ only and use trigger values $\epsilon_{\rho}$
of $0.02$ and $0.04$ while we set $\alpha$= $0.7$. The ratio of grid
spacings of refinement level $l$ and the base level permits a
level-dependent refinement via the choice of the exponent $\xi$. Here
we fix $\xi=0.6$, which makes refinements on higher levels
successively easier. Finally, we set the additional filter value
$\epsilon=0.01$ to prevent the refinement of small ripples.

\begin{figure}
  \center\includegraphics[width=\columnwidth]{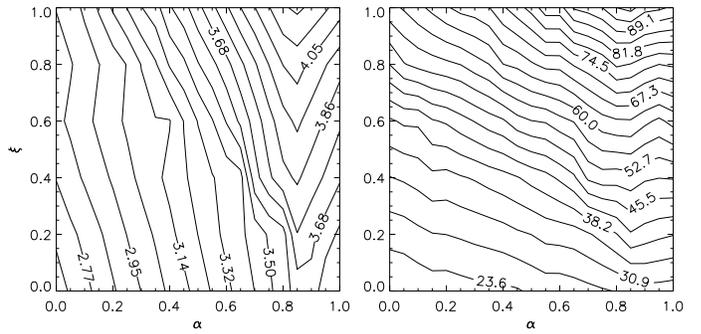}
  \caption{Effect of the adjustable parameters $\alpha$ and $\xi$ on
   the accuracy/efficiency of the mesh refinement. Contour lines
   indicate the maximum relative error (in percent, left panel), and
   the corresponding AMR speed-up (right panel) for the case of
   $\epsilon_{\rho}=0.02$ and with respect to the fully resolved
   reference run.}
  \label{fig:amr_param}
\end{figure}

While this general form of the refinement criterion is very flexible,
the large number of free parameters renders the optimal adjustment to
a particular problem a tedious task requiring a certain amount of
experience. Moreover, since AMR always implies a trade-off between
accuracy and efficiency, the {\it optimal} choice is by no means
clear-cut. This is illustrated in Fig.~\ref{fig:amr_param}, where we
show the variation of the accuracy and AMR speed-up\footnote{For a
description of the test case, see Sec.~\ref{sec:den3} below.} as a
function of the two supplementary parameters $\alpha$ and $\xi$ of the
classical refinement estimator (\ref{eq:classical}). One can, in
principle, use these graphs to find the parameter set which yields the
highest execution speed at a given level of fidelity. However, this
procedure is limited to simple test cases, and its universality cannot
easily be proofed.

\subsection{Refinement based on the Field condition}

In an alternative approach particularly tailored to trace condensation
interfaces, we apply grid refinement wherever the local Field-length
$\lF$ is not resolved by at least three grid cells. For fine tuning,
this number could, furthermore, be varied as a linear function of the
refinement level -- although we do not make use of this dispensable
feature here. The main advantage of the proposed new method lies in
the fact that it is virtually parameter free. This implies that the
intricate determination of suitable values for $\alpha$ and $\xi$ (as
described above) can be avoided.

The evaluation of $\lF$, given by Eq.~(\ref{eq:lF}), is
straightforward since the values of $\kappa$ and $T$ have already been
computed for the heat flux term. To minimise overheads, they are
stored in auxiliary arrays. It is evident that the definition of the
Field-length requires the inclusion of a heat conduction term in the
energy equation. Contrary to this, the FLASH code supports refinement
based on the local cooling time $\tau=\varepsilon/\rho^2\!\Lambda$,
which is independent of $\kappa$. If, in our case, $\kappa$ is scaled
with $\rho$ (see below), the two methods should perform similarly.


\section{Results}

\subsection{Gaussian density perturbation} \label{sec:den3}

\begin{figure}
  \center\includegraphics[width=0.9\columnwidth,%
                          bb=15 8 505 342,clip]{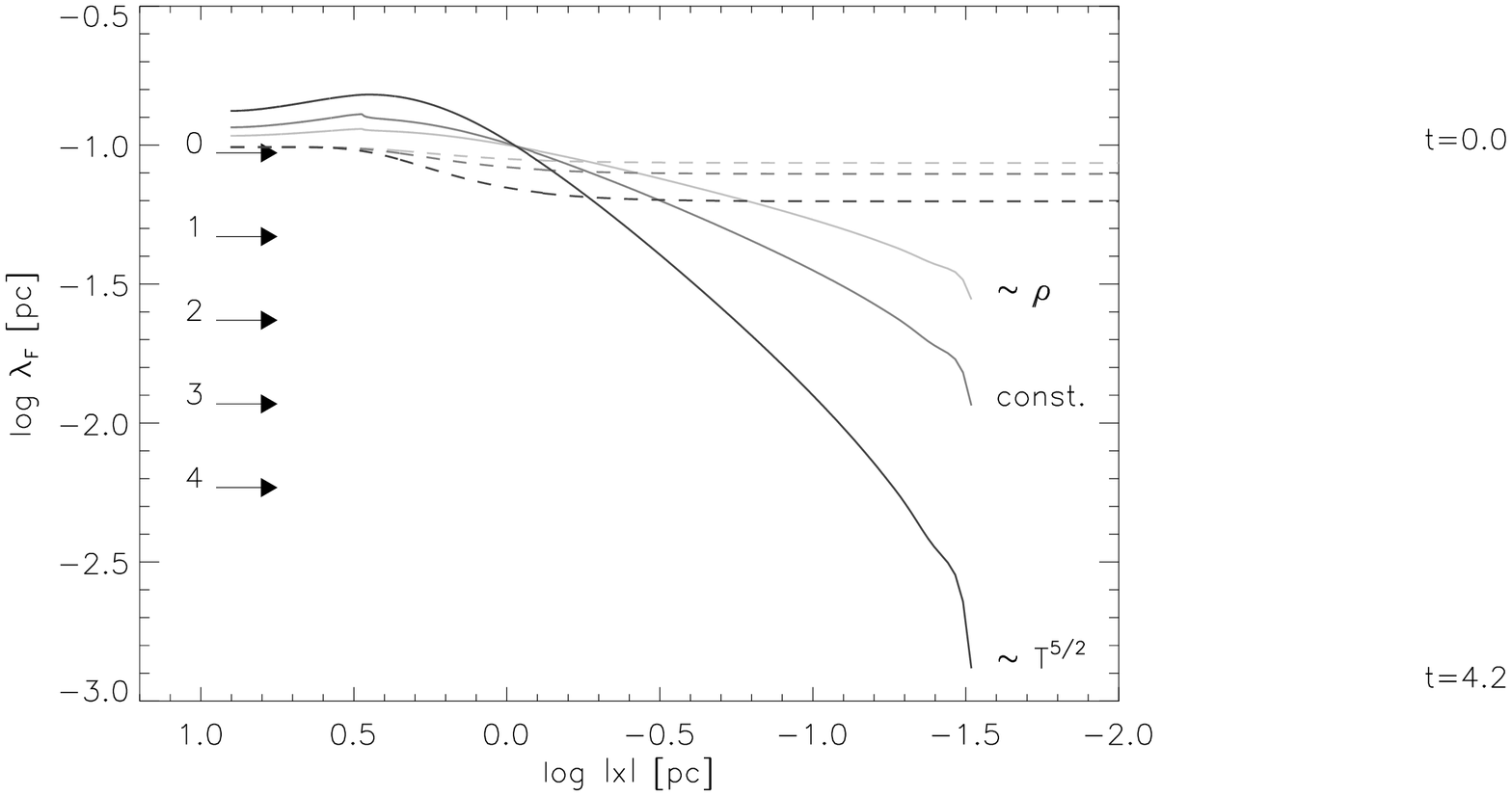}
  \caption{Field-length as a function of radius for the problem DEN3
    described in \citet{2002ApJ...577..768S} for the initial profile
    (dashed lines), and the final profile at $t=4.2\Myr$ (solid
    lines). The diffusion is set at a level of $1.02\times
    10^4\erg\,\cm^{-1}\K^{-1}\s^{-1}$ and scaled as indicated by the
    labels. Arrows mark a spacing of $3\Delta x$ at different
    refinement levels.}
  \label{fig:field_length}
\end{figure}

As a test problem, we revisit the 1D case of a density perturbation as
considered in the model DEN3 from \citet{2002ApJ...577..768S},
hereafter SSVSG. In this setup, the gas with density $n=1\cmt$
initially rests at its equilibrium pressure of $p/k_{\rm
B}=2500\K\cmt$ and is perturbed by a Gaussian overdensity of amplitude
$0.2$, FWHM of $3\pc$, and centred within a domain of $16\pc$. Unlike
in our simulations, SSVSG do not apply thermal conduction. This means
that we have to choose high enough resolution, such that the
conduction (which mainly becomes important near the grid scale) does
not overly affect the features seen in their solution. As a reference,
we perform a uni-grid run at a linear resolution of 8192 cells,
corresponding to a grid spacing of $\simeq 2\times10^{-3}\pc$. The
actual level of conduction is set such that $\lF$ is resolved properly
in the reference run. This yields a value for $\kappa$ which is about
a factor of 500 larger than the physical Spitzer value in the limit of
vanishing electron density. The thus even smaller structures expected
in a physically meaningful scenario illustrate the need for applying
multiscale techniques.

Fig.~\ref{fig:field_length} shows a plot of the Field-length for the
initial and final density and temperature profiles, where we use three
different prescriptions for the conduction: Scaling $\kappa$ with
$\rho$ is typically used together with constant kinematic viscosity
yielding a constant value for the Prandtl-number along with a constant
numerical time step, which is computationally favourable. This is also
reflected in the fact that the Field-length only shows a weak
variation with density. For the case of constant $\kappa$, this
variation is stronger and accounts for about one order of magnitude in
the problem under consideration. Finally, for a Spitzer-type
conductivity with $\kappa\sim T^{5/2}$, we obtain a variation of over
two orders of magnitude in $\lF$ rendering this case particularly
interesting for AMR. In the following, we restrict ourselves to the
case of a constant coefficient $\kappa=1.02\times 10^4
\erg\cm^{-1}\K^{-1}\s^{-1}$.

\begin{figure*}
  \center\includegraphics[width=1.80\columnwidth]{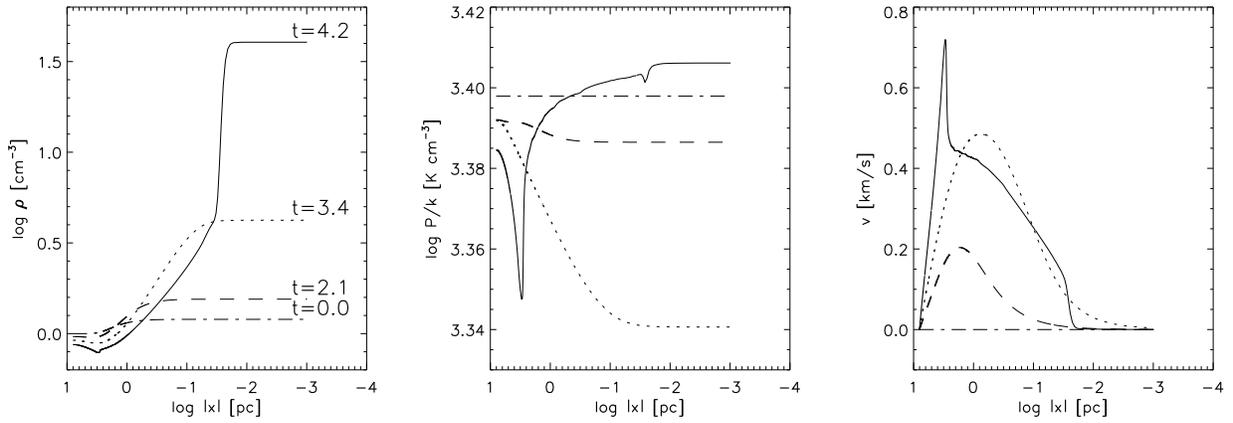}
  \caption{Temporal evolution of density (left), thermal pressure
    (centre), and velocity (right), for our 512+4 run with constant
    $\kappa$ and refinement based on the Field condition \citep[same
    as Fig.~2 of][]{2002ApJ...577..768S}.}
  \label{fig:ssvsg2002_fig2}
\end{figure*}

In general, the results are only marginally affected when conduction
is included and agree well with the solution of SSVSG. In
Fig.~\ref{fig:ssvsg2002_fig2}, the resulting profiles for the run with
a 512 cell base-grid plus 4 levels of refinement (based on the Field
condition) are plotted. Before we proceed with an analysis of the AMR
performance, we want to compare our reference run 8192+0 (which is
barely distinguishable from the one shown in
Fig.~\ref{fig:ssvsg2002_fig2}) with the profiles from SSVSG: The most
noticeable discrepancy is in the overall level of the thermal
pressure. Due to the assumed periodicity at the domain boundaries, the
mean pressure becomes a highly fluctuating quantity (cf. central panel
of Fig.~\ref{fig:ssvsg2002_fig2}). Apart from the different offsets,
the pressure profiles reasonably match. Notably, there is a small
pressure dip at the interface of the condensed structure in our runs
that is not seen in SSVSG. A similar feature is observed in panel (d)
of their Fig.~7 (model DEN75), although much spikier. This leaves the
possibility that the feature is not resolved by the plot for model
DEN3 in Fig.~2 of SSVSG. A comparison run with $\kappa=0$ in fact
shows that, in our setup, the dip is broadened by the thermal
conduction and remains much narrower otherwise. Later investigations
on the topic consistently reveal a similar region of lower pressure
\citep[see e.g.][]{2003LNP...614..213V}, which can be attributed to
the fact that the pressure decreases with density in the unstable
regime. In this respect, the pressure dip reflects the S-shape of the
equilibrium cooling curve.

\begin{figure}
  \center\includegraphics[width=0.8\columnwidth]{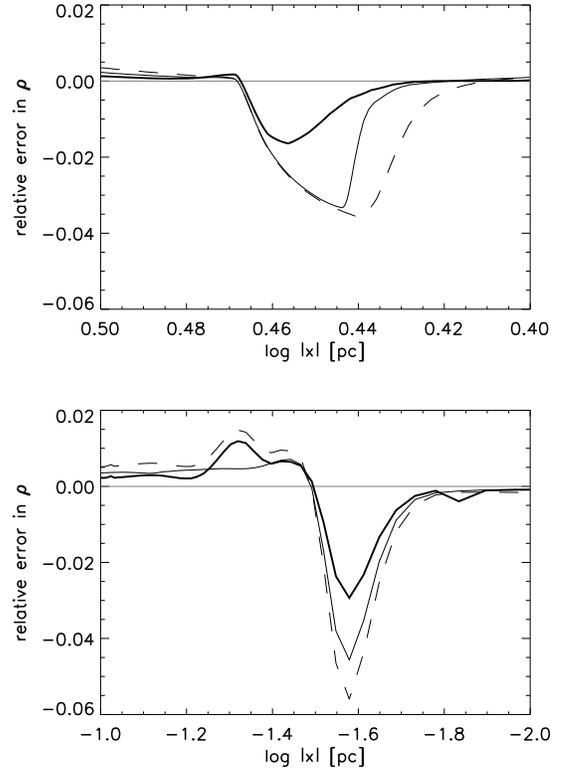}
  \caption{Relative errors of the density profiles at time $t=4.2\Myr$
    with respect to the fully resolved reference solution. The two
    intervals with the highest deviations are shown for refinement
    based on $\lF$ (thick line), $\varepsilon_{\rho}=0.02$ (solid
    line), and $\varepsilon_{\rho}=0.04$ (dashed line).}
  \label{fig:refine}
\end{figure}

In Fig.~\ref{fig:refine}, we plot the relative errors of three AMR
runs with respect to the fully resolved solution. The two panels
correspond to the two intervals with the strongest deviations, i.e.,
an outgoing wave (upper panel) and the condensation interface of the
cloud (lower panel). We compare the two runs based on the density
gradient criterion (with $\epsilon_{\rho}= 0.02$ and $0.04$) with the
run based on the Field condition. At the time $t=4.2\Myr$ the number
of AMR blocks for these runs are $820$, $476$, and $504$,
respectively. Although the number of refined cells is comparable, the
Field condition yields a relative error that is lower by a factor of
two at the condensation front. For $\epsilon_{\rho}=0.04$ ($0.02$) the
outgoing wave is resolved at $l=1$ ($3$), while the Field condition,
expectedly, is insensitive to this feature and does not refine
it. Still the relative error is lowest in this case, indicating that
the wave might be overly steepened by the refinement based on the
density gradients. This finding is rather surprising and deserves to
be looked at in greater detail.


\section{Discussion} \label{discussion}

\begin{figure}
  \center\includegraphics[width=0.8\columnwidth,bb=10 30 145 145,clip]{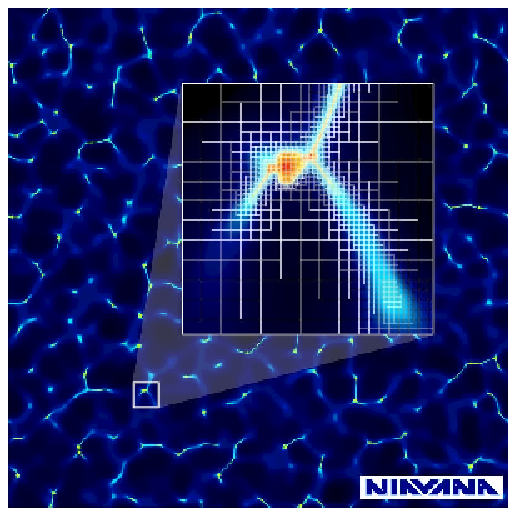}
  \caption{The Field-length criterion in action: development of the
    thermal instability from random fluctuations with Spitzer-type
    conductivity. The 2D simulation has a base grid of $256^2$ with
    refinement of up to 8 levels, covering scales from $0.003$ --
    $200\pc$.}
  \label{fig:ism}
\end{figure}

As is illustrated in Fig.~\ref{fig:ism}, thermal fragmentation
produces turbulent and extremely filamentary structures. Since
turbulence can only be modelled properly in three dimensions, adaptive
mesh refinement becomes highly beneficial, if not mandatory. Because
classical refinement strategies reach their limits if turbulence is
involved \citep[see e.g.][for a vorticity-based refinement
criterion]{2008MNRAS.388.1079I}, one has to seek for new tracers of
structural properties. This is particularly true in the case of a
thermally unstable medium.

Up to now, the criterion introduced by \citet{2004ApJ...602L..25K} is
widely disregarded by many members of the numerical astrophysics
community. Notable exceptions are a TI study by
\citet{2007ApJ...654..945B} and the MRI simulations of
\citet{2004ApJ...601..905P}. The opposite standpoint, represented by a
number of authors \citep[see
e.g.][]{2005ApJ...630..911G,2006ApJ...653.1266J}, is to neglect
thermal conduction altogether. The general argumentation is that
numerical diffusion defines a ``numerical Field-length'' that is
thought to sufficiently damp small-scale modes of the instability.
\citet{2004A&A...425..899D}, on the other hand, argue that the
microscopic heat conduction is suppressed perpendicular to the
magnetic field lines (and thus also isotropically for sufficiently
tangled fields) and that turbulent transport takes an important
role. Although this is certainly true, it does not aid the discussion
of whether explicit conduction should be included.

In our own simulations, we only observe artificial growth of modes
near the resolution limit in situations where the Courant number is
chosen inappropriately high or the gradient based refinement criterion
is improperly adjusted. The absence of unphysical growth at high
wavenumbers might thus, in fact, be attributed to the finite level of
numerical dissipation, which is never really negligible in
three-dimensional simulations.

In this respect, the Field criterion of
\citeauthor{2004ApJ...602L..25K} indeed seems of academic interest
only. Apart from its actual necessity, we, however, demonstrate that
the very condition can successfully be used as a refinement criterion
for adaptive mesh simulations of the interstellar medium -- an
approach that is very similar to the refinement procedure based on the
Jeans-length as introduced by \citet{1997ApJ...489L.179T} for
self-gravitating clouds. In a simple 1D test case, the new refinement
estimator is found to produce more accurate results at comparable
numerical cost than conventional criteria.

The overhead of evaluating $\lF$ is low compared to the computation of
the numerical fluxes. Within NIRVANA, the determination of the heat
flux already requires the evaluation of the gas temperature and the
(non-uniform) conductivity coefficient. These fields can therefore be
stored temporarily such that the only additional expense comes from
evaluating the cooling function.

The definition of $\lF$, nevertheless, implies the inclusion of
thermal conduction, which is not a standard feature in many available
codes. An alternative approach (implemented in the FLASH code) is to
use the cooling time instead, which is independent of $\kappa$. On the
other hand, there is evidence (Piontek, in prep.) that heat conduction
is in fact {\it physically} relevant for the formation of molecular
clouds, providing further motivation for the use of the proposed
scheme.

Compared to the classical mesh refinement based on gradients, which
has to be fine-tuned via multiple parameters according to different
situations, the Field criterion is virtually parameter-free and gives
appropriate results irrespective of the particular setup. If, of
course, other features like shock fronts are supposed to be adequately
resolved, one still has to rely on a combination with conventional
refinement criteria; this does, however, not impose any difficulties.

\begin{acknowledgements}
  This work used the NIRVANA code v3.3 developed by Udo Ziegler at the
  Astrophysical Institute Potsdam. We thank Enrique Vazquez and Simon
  Glover for the discussion related to this work and acknowledge the
  helpful comments by the anonymous referee.
\end{acknowledgements}


\end{document}